\documentclass[fleqn,twoside]{article}
\usepackage{amsmath}  % for align,
\usepackage{amssymb} % for \lesssim
\usepackage{espcrc2} % for elsevier stuff

\usepackage{graphicx}
\usepackage{psfrag}

\newcommand{\be}{\begin{equation}}
\newcommand{\ee}{\end{equation}}

\title{String breaking with Wilson loops?}

\author{Slavo Kratochvila\address[ETHZ]{Institute for Theoretical Physics, ETH Z\"{u}rich, CH-8093 Z\"{u}rich, Switzerland}\thanks{Talk presented by S. Kratochvila} and
        Philippe de
        Forcrand\addressmark[ETHZ]\address[CERN]{Theory Division, CERN, CH-1211 Geneva 23, Switzerland}}

\begin{document}

\begin{abstract}

A convincing, uncontroversial observation of string breaking, when
the static potential is extracted from Wilson loops only, is still
missing. This failure can be understood if the overlap of the
Wilson loop with the broken string is exponentially small. In that
case, the broken string ground state will only be seen if the
Wilson loop is long enough. Our preliminary results show string
breaking in the context of the $3d$ $SU(2)$ adjoint static
potential, using the L\"uscher-Weisz exponential variance
reduction approach. As a by-product, we measure the fundamental
$SU(2)$ static potential with improved accuracy and see clear
deviations from Casimir scaling.

\vspace{1pc}
\end{abstract}

\maketitle

\section{INTRODUCTION}
The breaking of a long flux tube between two quarks into a
matter-antimatter pair is one of the most fundamental phenomena in
QCD. Three approaches have been used to measure the static
potential: (i) Correlation of Polyakov Loops, at finite T
\cite{Laermann:1998gm}. (ii) Variational Ansatz using two types of
operators: One for the string-like state and one for the broken
string state
\cite{Philipsen:1998de,Stephenson:1999kh}. (iii) Wilson loops \cite{Poulis:1995nn,Aoki:1998sb}. \\
\vspace{-0.4cm}

String breaking has been seen using the first two methods, but no
clear signal using the third one has been observed. The failure of
the Wilson loop method seems to be mainly due to the poor overlap
of the operators with the broken state. The strong coupling model
\cite{Drummond:1998ar} uses the heavy quark expansion to show that
there is an exponentially small overlap. Based on a topological
argument, \cite{Gliozzi:1999cv} suggests that there may be no
overlap at all. We show that there is a small overlap but not
nearly as small as predicted.

\section{STATIC POTENTIAL}

The static potential helps to characterize confining forces.
Assume that a static charge and a static anticharge are separated
by a distance $R < R_b$, where $R_b$ is the string breaking
distance. If $R$ is increased to a value larger than $R_b$, a
matter-antimatter pair will be created which screens the static
charges. A further increase of $R$ has no effect on the static
potential - it remains $\approx$ constant, this indicates string
breaking. In our case, we deal with adjoint static charges which
are screened by the gluons of
the gauge field. The object adjoint\_charge-gluon is called a gluelump. \\
\vspace{-0.4cm}

In a Hamiltonian formulation, a static charge and a static
anticharge are created at Euclidean time 0, forming a state $|\phi
\rangle$, and annihilated at time $T$. This can be expressed as
the Wilson loop $W(R,T)$: \be\label{eq:wilson_loop} \langle \phi |
e^{-T H} | \phi \rangle = W(R,T) \ee On the other hand, one can
expand the left side in (\ref{eq:wilson_loop}) in the eigenbasis
$\Psi^{(n)}$ of the Hamiltonian. At large $T$, only the ground
state survives:
\begin{align}\label{eq:eigenbasis}
\langle \phi | e^{-T H} | \phi \rangle & = \sum_{n=0}^{\infty} |
\langle \Psi^{(n)} | \phi \rangle |^{2} e^{-T E_n}  \notag \\
& \underset{T \rightarrow \infty}{\sim }  |
\langle \Psi^{(0)} | \phi \rangle |^{2} e^{-T V(R)}
\end{align}

By using (\ref{eq:wilson_loop}) and (\ref{eq:eigenbasis}), the static potential $V(R)$,
which is defined as the energy of the ground state $E_0$, can be connected with
the Wilson loop:
\be \label{eq:VR}
V(R) = - \underset{T \rightarrow
\infty}{\mathrm{lim}} \frac{1}{T} \log W(R,T)
\ee
This holds as long as the overlap  $\langle  \phi | \Psi^{(0)}\rangle$ does not vanish. \\
In section \ref{sec:results}, we fit the energy of the unbroken
case using the Ansatz:
\be \label{eq:VR_Ansatz}
V(R) \sim V_0
-\frac{c}{R}+ \sigma R
\ee where $\sigma$ is the string tension.

\section{METHOD}
We are using a $3d$-lattice with extent $48^2 \times 64$ at $\beta=\frac{4}{a g^2}=6.0$. The gauge group is $SU(2)$. We
measure adjoint static charges in order to avoid costly dynamical quarks.
The price to pay is that the signal decays much faster because $V_{adj} \approx \frac{8}{3} V_{fund}$ (Casimir scaling). \\
\vspace{-0.4cm}

We make use of L\"uscher-Weisz exponential variance reduction
\cite{Luscher:2001up}. This method generalizes Multihit
\cite{Parisi:1983} from single links to link-link-correlators
$\textbf{T}(n a)$.
% Those are
%tensors and by defining an appropriate tensor-multiplication, we can preserve the path-order of the Wilson loop.
The possibility to use a hierarchical scheme allows us to measure long Wilson loops up to the desired accuracy.
Using the notation introduced in \cite{Luscher:2001up}, we actually implement the following scheme:\\
\begin{equation*}
W(R,T) = \textbf{L}(0)_{\alpha\gamma} \{ \cdots  \biggl[\Bigl[  [ \textbf{T}(n a)  ][  \textbf{T}( (n+1)a) ]  \Bigr] \cdot
\end{equation*}
\be
 \Bigl[
 [  \textbf{T}((n+2)a)  ][  \textbf{T}((n+3)a) ]  \Bigr]  \biggr] \cdots  \}_{\alpha\beta\gamma\delta} \textbf{L}(T)^{*}_{\beta\delta}
\ee
The parameters are the following: The innermost average $[
\textbf{T}(n a)]$ was calculated from 10 sets of time-like links,
each obtained after 10 updates (1HB/4OR). The second-level average
$\Bigl[ [ \textbf{T}(n a)  ][ \textbf{T}((n+1)a) ] \Bigr]$ was
calculated from 160 averages of $[ \textbf{T}(n a)]$ separated by
200 updates of the spatial links on the time-slice $[(n+1) a]$.
Finally, the third-level average was calculated from 39
second-level averages separated by 200 updates of the spatial
links on the
time-slice $[(n+2) a]$. \\
\vspace{-0.4cm}

To reduce contributions from excited states $\Psi^{(n \neq 0)}$,
we replace the simple spatial transporters
$\textbf{L}(0)_{\alpha\gamma}$ and
$\textbf{L}(T)^{*}_{\beta\delta}$ with "staples", which are
constructed in the following way: (i) After each calculation of
second level averages,
\includegraphics[height=0.25cm]{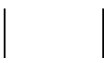} , we form
the smeared adjoint spatial links,
\includegraphics[height=0.075cm]{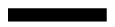} , at the
time-slice $(n+2)a$. (ii) We multiply them with the second level
averages to obtain the staples
\includegraphics[height=0.25cm]{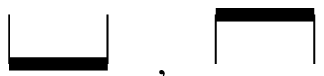}.
(iii) This procedure is repeated 39 times to provide an error reduction in the new spatial transporters. \\

After processing 40 configurations as above, we obtain the
following preliminary results.

\section{RESULTS}\label{sec:results}

We measure both the adjoint and the fundamental static potential
(Fig. 1). The spatial separation $R$ starts at 2 and is increased
up to 12. In the adjoint case, the Ansatz (\ref{eq:VR_Ansatz})
works well for the energy $V'(R)$ of the unbroken string in the
whole range of $R$. However, at $R \geq 10$ the minimal energy
remains constant, $V(R \geq 10)=1.95(10)$.
Therefore, we observe string breaking at a distance $R_b \approx 10$. \\
\vspace{-0.4cm}

The dotted line at 2.06(1) represents twice the energy of a
gluelump using a direct measurement \cite{Poulis:1995nn} on our
lattice. This value is close to the groundstate potential for
$R>R_b$ (See Fig.1). Both $R_b \approx 10$ \cite{Stephenson:1999kh}, as the string breaking
distance, and $2 E(Qg)$ \cite{Stephenson:1999kh,Poulis:1995nn}, as twice the gluelump-energy,
are in agreement with the literature.\\
\vspace{-0.4cm}

%(papers using the variational approach \cite{Stephenson:1999kh})(
%and with papers measuring the gluelump-mass
%\cite{Stephenson:1999kh}\cite{Poulis:1995nn}). \\
In the fundamental case, the Ansatz (\ref{eq:VR_Ansatz}) again
works very well. An important issue is the fulfilment of the
Casimir scaling law. At $R \lesssim 4$, $\frac{V_{adj}}{V_{fund}}
\approx \frac{8}{3}$ holds as required by perturbation theory.
However at large distances, the slopes of the potentials are in
the ratio $2.31(1) < \frac{8}{3}$.\vspace{-0.5cm}

\begin{center}
%\footnotesize
%\textbf{Static Potential}
%\normalsize
\psfrag{EQgLABEL}{\tiny 2E(Q$\bar{\mathrm{g}}$)}
\psfrag{ADJPOTEN}{\tiny V$_{\mathrm{adj}}$}
\psfrag{SCFUPOTE}{\tiny $\frac{8}{3}$V$_{\mathrm{fund}}$}
\psfrag{VPRLABEL}{\tiny V'(R)}
\psfrag{V(R)}{\tiny V(R)}
\psfrag{R}{\tiny R}
\includegraphics[height=7.5cm,angle=-90]{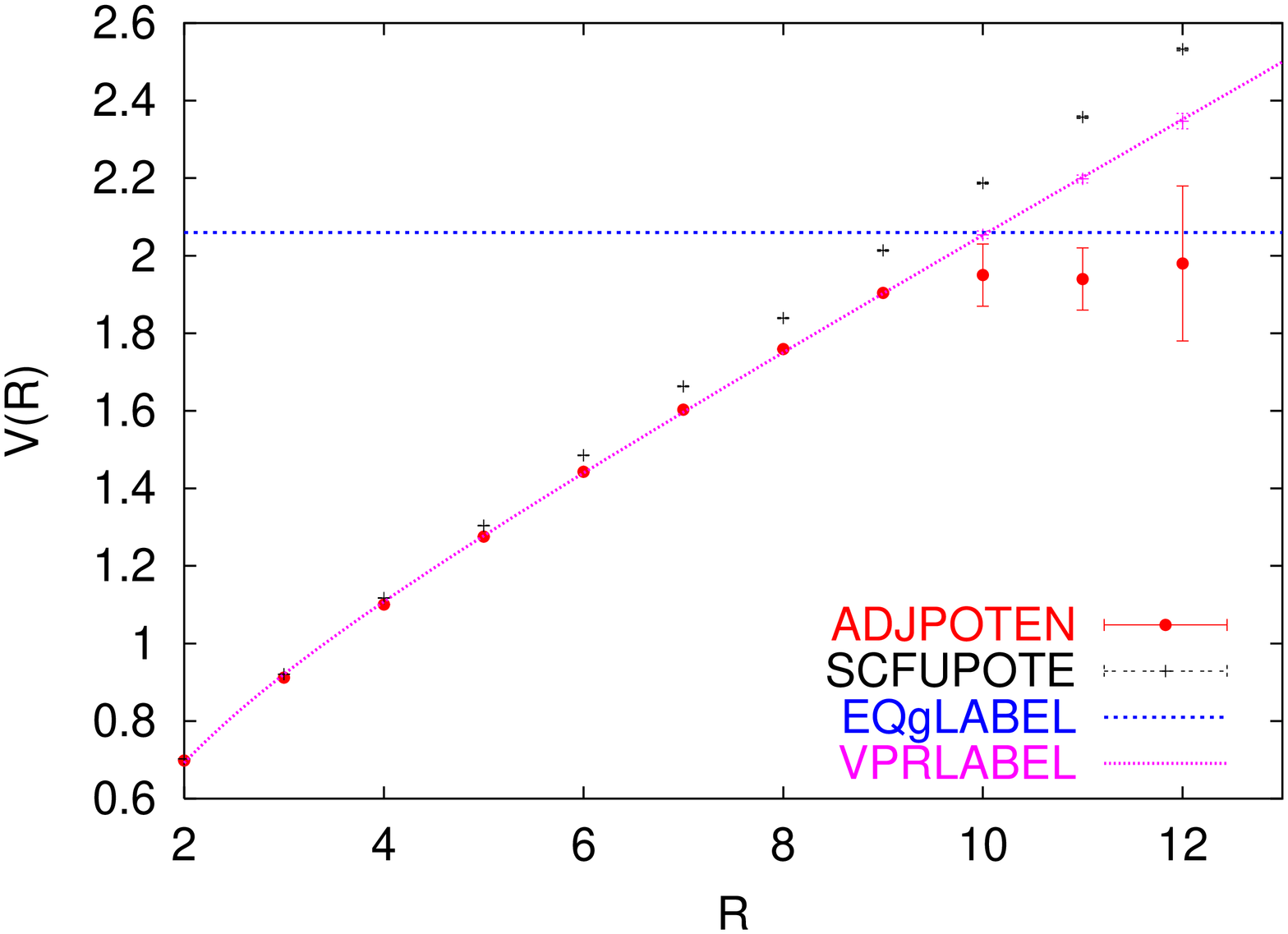}
\end{center}
\vspace{-0.2cm} \footnotesize Fig. 1. The adjoint and
$\frac{8}{3}$ fundamental static potentials $V(R)$ vs $R$. The
horizontal line at 2.06(1) represents
twice the energy of a gluelump. \normalsize \\

The value of the static potential $V(R)$ at a fixed $R$, e.g.
$R=12$, is determined by the slope of the logarithm of the Wilson
loop (\ref{eq:VR}). In Fig. 2 it can be seen that a single
exponential fit is not sufficient in the adjoint case. At small
temporal extent $T$ we get a larger slope than at large $T$. This
can be explained as follows: At small $T$ the signal is dominated
by the unbroken string state. The broken string state can only be
observed after $T$ is large.
%\begin{center}
%\footnotesize
%\textbf{Single-exponential fits: $c e^{-V T}$}
%\normalsize
%\includegraphics[height=7.5cm,angle=-90]{./single_exp_fit.eps}
%\end{center}
%\footnotesize
%Fig. 2. Wilson loop $W(R,T)$ vs $T$, $R=12$ fixed. Single-exponential fits do not match
%all data points. The unbroken string state can be fitted by using data from $T=4$ up to 12, the broken
%string state by using data from $T=12$ up to $T=28$. \normalsize \\
The way to fit all data points is to use the double-exponential
fit ${\sim c_0 e^{-V'(R) T} + c_1 e^{-2 E(Q \bar{g}) T}}$. $V'(R)$
corresponds to the energy of the unbroken string, $E(Q \bar{g})$
to the energy of the gluelump. The ratio of the amplitudes is
$\frac{c_0}{c_1} \sim 10^4$, the turning point is at T$_t \approx
20$.\vspace{-0.5cm}

\begin{center}
%\footnotesize
%\textbf{Double-exponential fit $\sim c_0 e^{-V'(R) T} + c_1 e^{-2 E(Q \bar{g}) T}$}
%\normalsize
\psfrag{tp}{\tiny T$_t$} \psfrag{turpoint}{\tiny T$_t \approx 20$}
\psfrag{c0c1}{\tiny $\frac{c_0}{c_1} \sim 10^4$}
\includegraphics[height=7.5cm,angle=-90]{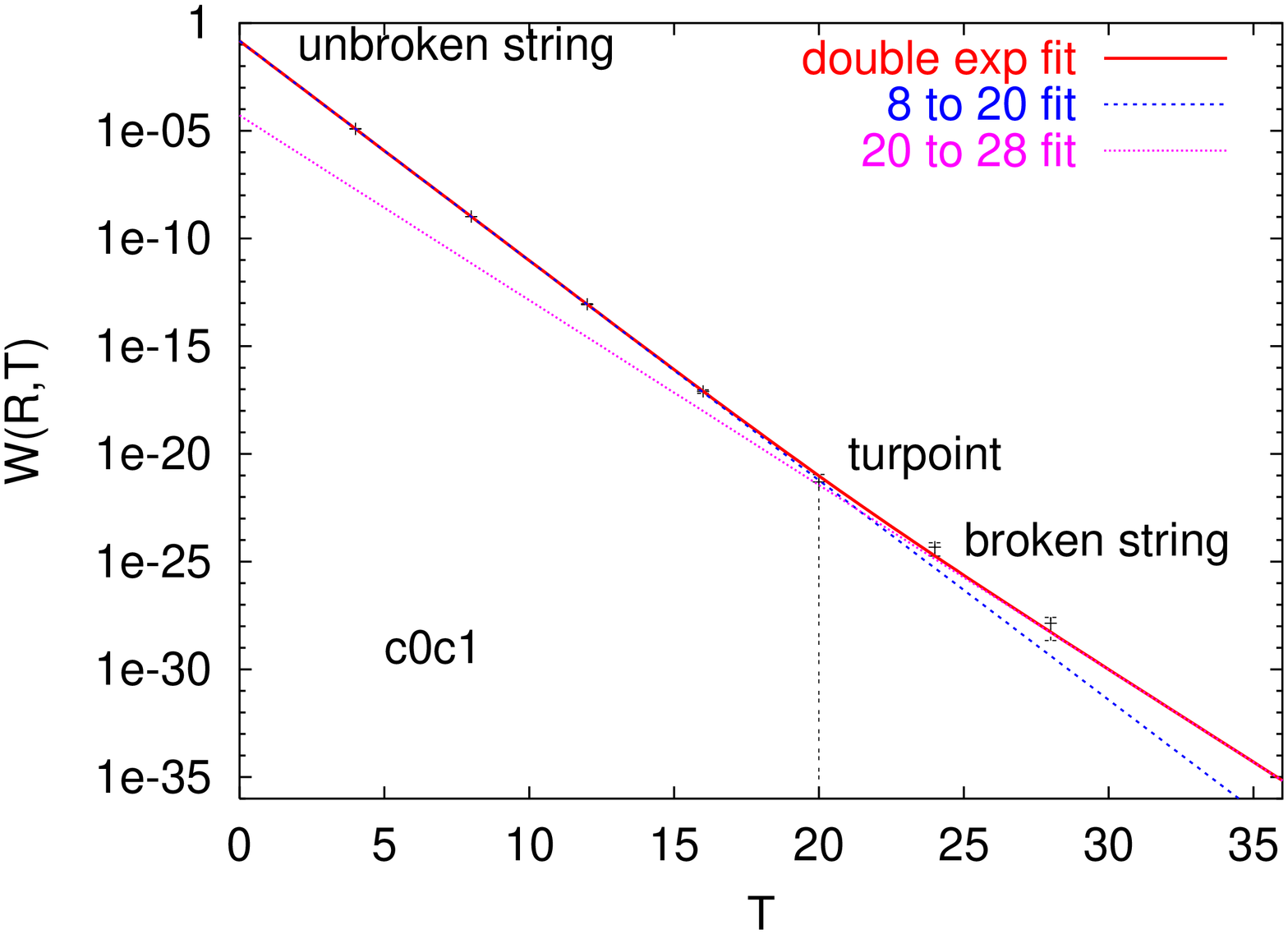}
\end{center}
\vspace{-0.2cm} \footnotesize Fig. 2. Adjoint Wilson loop $W(R,T)$
vs $T$, for $R=12$. The double-exponential fit matches all data
points. Single-exponential fits (dotted lines) do not. The
unbroken string state can be fitted by using data from $T=8$ up to
20, the broken string state by using data from $T=20$ up to 28.
Note how small a signal can be measured.
 \normalsize

\section{CONCLUSION}
In the context of the $3d$ $SU(2)$ adjoint static potential,
string breaking has been observed using Wilson loops only. \\
\vspace{-0.4cm}

It is a necessary condition that both $R$ and $T$ have to be
larger than the string breaking distance $R_b$.
Note, however, that the variational approach allows to bypass this requirement. \\
\vspace{-0.4cm}

The strong coupling model \cite{Drummond:1998ar} is too
pessimistic. A necessary condition to be able to measure the
broken state is that the signal should be larger than the one from
the unbroken state: ${c_1 e^{-2 M_{Qg} T} \gg c_0 e^{- V(R) T}}$.
The heavy quark expansion results in the ratio $\frac{c_0}{c_1}
\sim e^{M_{Qg} R} $, which would imply ${T \gg 43}$ at ${R=12}$,
whereas in our simulation the turning point was {T$_t \approx 20$}. \\
\vspace{-0.4cm}

% On the whole, it is difficult but feasible to extract string breaking within full QCD
A similar observation using Wilson loops in full QCD should be
difficult but feasible \cite{Trottier:2002ry}.


\begin{thebibliography}{9}
%\cite{Laermann:1998gm}
\bibitem{Laermann:1998gm}
E.~Laermann, C.~DeTar, O.~Kaczmarek and F.~Karsch,
%``String breaking in lattice QCD,''
Nucl.\ Phys.\ Proc.\ Suppl.\  {\bf 73} (1999) 447;
%%CITATION = HEP-LAT 9809105;%%
C.~DeTar, U.~M.~Heller and P.~Lacock,
%``First signs for string breaking in two-flavor QCD,''
Nucl.\ Phys.\ Proc.\ Suppl.\  {\bf 83} (2000) 310;
%%CITATION = HEP-LAT 9909078

%\cite{Philipsen:1998de}
\bibitem{Philipsen:1998de}
O.~Philipsen and H.~Wittig,
%``String breaking in non-Abelian gauge theories with fundamental matter  fields,''
Phys.\ Rev.\ Lett.\  {\bf 81} (1998) 4056 [Erratum-ibid.\  {\bf
83} (1999) 2684];
%%CITATION = HEP-LAT 9807020;%%
F.~Knechtli and R.~Sommer  [ALPHA collaboration],
%``String breaking in SU(2) gauge theory with scalar matter fields,''
Phys.\ Lett.\ B {\bf 440} (1998) 345;
%%CITATION = HEP-LAT 9807022;%%
O.~Philipsen and H.~Wittig,
%``String breaking in SU(2) Yang-Mills theory with adjoint sources,''
Phys.\ Lett.\ B {\bf 451} (1999) 146;
%%CITATION = HEP-LAT 9902003;%%
P.~de Forcrand and O.~Philipsen,
%``Adjoint string breaking in 4d SU(2) Yang-Mills theory,''
Phys.\ Lett.\ B {\bf 475} (2000) 280;
%%CITATION = HEP-LAT 9912050;%%
K.~Kallio and H.~D.~Trottier,
%``Adjoint 'quarks' on coarse anisotropic lattices: Implications for  string breaking in full QCD,''
arXiv:hep-lat/0001020;
%%CITATION = HEP-LAT 0001020;%%
P.~Pennanen and C.~Michael  [UKQCD Collaboration],
%``String breaking in zero-temperature lattice QCD,''
arXiv:hep-lat/0001015;
%%CITATION = HEP-LAT 0001015;%%
C.~W.~Bernard {\it et al.}  [MILC Collaboration],
%``Zero temperature string breaking with staggered quarks,''
Nucl.\ Phys.\ Proc.\ Suppl.\  {\bf 94} (2001) 546.
%%CITATION = HEP-LAT 0010066;%%

%\cite{Stephenson:1999kh}
\bibitem{Stephenson:1999kh}
P.~W.~Stephenson,
%``Breaking of the adjoint string in 2+1 dimensions,''
Nucl.\ Phys.\ B {\bf 550} (1999) 427.
%%CITATION = HEP-LAT 9902002;%%

%\cite{Poulis:1995nn}
\bibitem{Poulis:1995nn}
G.~I.~Poulis and H.~D.~Trottier,
%``'gluelump' spectrum and adjoint source potential in lattice QCD in three-dimensions,''
Phys.\ Lett.\ B {\bf 400} (1997) 358.
%%CITATION = HEP-LAT 9504015;%%

%\cite{Aoki:1998sb}
\bibitem{Aoki:1998sb}
S.~Aoki {\it et al.}  [CP-PACS Collaboration],
%``The static quark potential in full QCD,''
Nucl.\ Phys.\ Proc.\ Suppl.\  {\bf 73} (1999) 216;
%%CITATION = HEP-LAT 9809185;%%
G.~S.~Bali {\it et al.}  [TXL Collaboration],
%``Static potentials and glueball masses from QCD simulations with
%Wilson  sea quarks,''
Phys.\ Rev.\ D {\bf 62} (2000) 054503;
%[arXiv:hep-lat/0003012].
%%CITATION = HEP-LAT 0003012;%%
B.~Bolder {\it et al.},
%``A high precision study of the Q anti-Q potential from Wilson loops in  the regime of string breaking,''
Phys.\ Rev.\ D {\bf 63} (2001) 074504.
%[arXiv:hep-lat/0005018].
%%CITATION = HEP-LAT 0005018;%%

%\cite{Drummond:1998ar}
\bibitem{Drummond:1998ar}
I.~T.~Drummond,
%``Strong coupling model for string breaking on the lattice,''
Phys.\ Lett.\ B {\bf 434} (1998) 92.
%%CITATION = HEP-LAT 9805012;%%

%\cite{Gliozzi:1999cv}
\bibitem{Gliozzi:1999cv}
F.~Gliozzi and P.~Provero,
%``When QCD strings can break,''
Nucl.\ Phys.\ Proc.\ Suppl.\  {\bf 83} (2000) 461.
%%CITATION = HEP-LAT 9907023;%%

%\cite{Luscher:2001up}
\bibitem{Luscher:2001up}
M.~Luscher and P.~Weisz,
%``Locality and exponential error reduction in numerical lattice gauge  theory,''
JHEP {\bf 0109} (2001) 010;
%%CITATION = HEP-LAT 0108014;%%
P.~Majumdar,
%``Experiences with the multi-level algorithm,''
arXiv:hep-lat/0208068.
%%CITATION = HEP-LAT 0208068;%%

%\cite{Parisi:1983}
\bibitem{Parisi:1983}
G.~Parisi, R.~Petronzio and F.~Rapuano, Phys.\ Lett.\ B {\bf 128}
(1983) 418.

%\cite{Trottier:2002ry}
\bibitem{Trottier:2002ry}
H.~D.~Trottier and K.~Y.~Wong,
%``Static potential and local color fields in unquenched three-dimensional  lattice QCD,''
arXiv:hep-lat/0209048, these proceedings.
%%CITATION = HEP-LAT 0209048;%%


\end{thebibliography}
\end{document}